\documentclass[10pt,twocolumn,preprintnumbers,amsmath,amssymb,nofootinbib,superscriptaddress,prd]{revtex4-2}
\usepackage[dvipsnames]{xcolor}
\usepackage{comment}
\usepackage{bm}
\usepackage{graphicx}
\usepackage{tabularx}
\usepackage[colorlinks,allcolors=RoyalBlue]{hyperref}
\usepackage[normalem]{ulem}
\usepackage{xspace}
\usepackage{hyperref,url}
\usepackage{orcidlink}

\newcommand{\dchisq}{\ensuremath{\Delta\chi^2}}

\newcommand{\lcdm}{$\Lambda$CDM\xspace}

\usepackage[capitalise]{cleveref}

\newcommand{\rev}[1]{{#1}}

\newcommand{\taumax}{\tau_{\rm max}}

\begin{document}
\title{Raising the reionization optical depth with inflationary  CMB  features }
\author{Tanisha Jhaveri}
\affiliation{Department of Astronomy and Astrophysics, University of Chicago, 5640 South Ellis Avenue, Chicago, IL, 60637, USA}
\affiliation{Kavli Institute for Cosmological Physics, Enrico Fermi Institute, University of Chicago, 5640 South Ellis Avenue, Chicago, IL, 60637, USA}
\author{Wayne Hu}
\affiliation{Department of Astronomy and Astrophysics, University of Chicago, 5640 South Ellis Avenue, Chicago, IL, 60637, USA}
\affiliation{Kavli Institute for Cosmological Physics, Enrico Fermi Institute, University of Chicago, 5640 South Ellis Avenue, Chicago, IL, 60637, USA}
\author{Vivian Miranda}
\affiliation{C. N. Yang Institute for Theoretical Physics, Stony Brook University, Stony Brook, NY 11794, USA}

\begin{abstract}
Within the highly successful $\Lambda$CDM paradigm established with cosmic microwave background (CMB) anisotropy measurements, the optical depth through reionization $\tau$ is  the most uncertain due both to the difficulty in measuring large-angle polarization and the assumptions made in their interpretation.  Currently, for the Planck primary data in the flat $\Lambda$CDM cosmology with slow-roll inflation and standard reionization, the one-sided 95\% upper limit for $\tau$ is  $\taumax=0.0696$. Yet when all current CMB measurements {\it excluding} large-angle polarization are combined with baryon acoustic oscillation (BAO) measurements, the one-sided 95\% lower limit is an incompatible $\tau_{\rm min}=0.074$.  
If the long-standing low-power feature of the temperature measurements is interpreted as physically originating from inflation then $\tau$ inferred from large-angle polarization becomes larger.  Marginalizing over templates of the low-power feature based on the generalized slow-roll formalism of inflation raises the Planck maximum to a more compatible $\taumax=0.075$ which further increases to $\taumax = 0.082$ with the inclusion of all CMB+BAO data.  
This marginalization does not assess the statistical significance of the low-power feature itself; rather, it shows that allowing a higher $\tau$ is a consequence of interpreting the anomaly as a physical  feature instead of a statistical fluctuation.
\end{abstract}

\date{\today}

\maketitle

\section{Introduction} 
\label{sec:intro}

Measurements of cosmic microwave background (CMB) anisotropy have been critical in establishing the $\Lambda$CDM  cosmology as a simple and enormously successful model whose 6 parameters correctly predict observed phenomena across a wide range of redshifts, scales and independent cosmological probes.

Of those 6 parameters, historically the weakest link has been $\tau$, the Thomson optical depth through reionization.  For the CMB, the optical depth is also the most difficult of the $\Lambda$CDM parameters to measure, since it requires an accurate characterization of the power in the very lowest multipoles of the CMB E-mode polarization (lowE). First detected by the Wilkinson Microwave Anisotropy Probe (WMAP) in cross-correlation with temperature, its value has evolved from
$\tau = 0.17 \pm 0.04$ \cite{WMAP:2003ggs} through improvements in measurements and analysis techniques to $\tau=0.089 \pm 0.014$ (WMAP9 \cite{WMAP:2012nax}),
$\tau=0.097 \pm 0.038$ (Planck 2013  \cite{Planck:2013pxb}),  
$\tau =0.079\pm 0.017$ 
(Planck 2015 \cite{Planck:2015fie}),  $\tau=0.0544 \pm 0.00755$ (Planck 2018 \cite{Planck:2018vyg}) and, with improved analysis after the final Planck release, reversing the downward trend to arrive finally at $\tau =0.059\pm 0.006$ \cite{Pagano:2019tci}.  Even within the Planck data, these final low values are a compromise between the even lower inferences from lowE and the indirect constraints from gravitational lensing due to its impact on the normalization parameter $A_s$ of the curvature power spectrum when $A_s e^{-2\tau}$ is fixed by the observations
(e.g.~\cite{Planck:2016tof,Planck:2018vyg}).

These low values of $\tau$ have more recently been a source of tension in $\Lambda$CDM with measurements of baryon acoustic oscillations (BAO) from the Dark Energy Spectroscopic Instrument (DESI)
\cite{DESI:2025zgx,Jhaveri:2025neg,Sailer:2025lxj}.  In standard $\Lambda$CDM, the BAO prefer a larger distance between the central BAO redshifts at $z\sim 1$ and recombination ($z\sim 1100$), which requires lowering the cold dark matter density $\Omega_c h^2$. However, the CMB does not fully accommodate this reduced value due to the sharpening of the acoustic peaks and lensing. Raising the optical depth to $\tau \sim 0.09$ offers a way for the CMB to accommodate these low $\Omega_c h^2$ effects.
Separately, the low optical depth measurements also impact the interpretation of the ubiquity of luminous ultra-high redshift galaxies (e.g. \cite{2024ApJ...969L...2F,2022ApJ...938L..15C,2023arXiv230602465E,2023ApJS..265....5H}). 

In this work, we study a hidden assumption in the lowE CMB inferences on $\tau$.  Its interpretation relies on an extrapolation of the inflationary curvature power spectrum from the well-measured small-scale CMB to that of the lowest multipoles.  This extrapolation allows the conversion of the measured lowE power and a known curvature source into an inference of $\tau$.
Moreover, since the first WMAP release, there has been a long-standing anomaly in the temperature power spectrum on these scales (lowT)  \cite{WMAP:2003ggs} that has been verified by independent observations from Planck
\cite{Planck:2019evm}.   Indeed the interpretation of the low power anomaly was in part responsible for the shifts in $\Lambda$CDM parameters between WMAP9 and Planck \cite{Planck:2016tof}, in particular the high values of $\Omega_c h^2$ that cause the tension with current BAO measurements.    

If the low power anomaly is interpreted as a physical effect of low power in the inflationary power spectrum instead of a statistical fluctuation in $\Lambda$CDM, then the natural consequence is that $\tau$ must increase to compensate and restore lowE power.   Indeed, that the lowE power exists at all requires that there be a finite floor on any cutoff in the curvature power spectrum on the horizon scale \cite{Mortonson:2009xk}.
This cutoff model also serves as a concrete proof of principle that the optical depth increases when the large scale curvature decreases \cite{Jhaveri:2025neg}, albeit at low statistical significance (see also \cite{Huang:2025xyf,Upadhyay:2026unf}).

Here we explore the maximum extent to which the optical depth can increase when considering the lowT power anomaly as originating from inflation.  Since the lowT power anomaly has remained consistent throughout datasets from WMAP to Planck, we employ fixed template forms for the curvature power spectrum from the generalized slow-roll (GSR) analysis of the Planck 2015 data \cite{Obied:2018qdr} and marginalize over its amplitude with the current CMB and BAO data.  The GSR templates consistently include the power spectrum oscillations that must be present when deviations from slow-roll are used to explain the anomalous features, while the fixed template forms avoid excessively overfitting statistical fluctuations in the data with too many parameters. 
The purpose of this work is not to assess the statistical significance of large-scale temperature features. Instead, we ask a conditional question: if the observed lowT feature is interpreted as a physical inflationary feature rather than a statistical fluctuation, what are the resulting implications for the maximal optical depth $\taumax$ and its role in CMB+BAO consistency?

The outline of the paper is as follows.  In Sec.~\ref{sec:GSR} we review the GSR approach to large scale features and in Sec.~\ref{sec:data} the datasets and methodology used.   We present results for raising the optical depth to the maximum allowed by Planck primary anisotropy in Sec.~\ref{sec:planck} and the further increase in inferred $\tau$ when BAO data are included in Sec.~\ref{sec:BAO}.
We discuss these results in Sec.~\ref{sec:discussion}.

\section{GSR Power Spectrum Features}
\label{sec:GSR}

The generalized slow-roll (GSR) formalism for large power spectrum features was developed by Ref.~\cite{Dvorkin:2009ne} extending earlier work \cite{Stewart:2001cd,Dodelson:2001sh,Choe:2004zg,Kadota:2005hv}. While  sufficient inflation requires the first slow-roll parameter to be small 
\begin{equation}
\epsilon_H \equiv - \frac{d\ln H}{d\ln a} \ll 1,
\end{equation}
it allows the second slow-roll parameter to satisfy
\begin{equation}
\left|\frac{d\ln \epsilon_H}{d\ln a}\right|\lesssim 1,
\end{equation}
as long as it is not large for a sufficient number of efolds that $\epsilon_H$ itself becomes large.
A transiently large second slow-roll parameter produces features in the inflationary curvature power spectrum that locally deviate from the usual power law predictions where 
\begin{equation}
\ln{\cal P} = \ln A_s + (n_s-1) \ln(k/k_0),
\end{equation}
and $k_0=0.05$\,Mpc$^{-1}$ by convention.  We will refer to this slow-roll case in the usual flat cold dark matter and cosmological constant model as ``$\Lambda$CDM."

The GSR formalism extends the validity of power spectrum predictions to ${\cal O}(1)$ transient deviations in this second slow-roll parameter. 
In particular, such deviations in the temporal history of $\epsilon_H$  are completely encapsulated in a generalization of the local tilt
\begin{equation}
G'(\ln s)=1-n_s + \delta G'(\ln s) ,
\end{equation}
where $s=\int d\ln a (c_s/aH)$ is the inflationary sound horizon measured from the end of inflation (see Ref.~\cite{Motohashi:2017gqb} for the explicit form in the effective field theory of inflation).
The power spectrum is then evaluated as an integral over this source function
\begin{align}
\ln {\cal P}(k) =
\int_{0}^\infty
\frac{ds}{s} W(ks) G'(\ln s)
{}& + \ln\left[ 1+ I_1^2(k)\right],
\end{align}
where
\begin{equation}
I_1(k) = \frac{1}{\sqrt{2}} \int_0^\infty \frac{ds}{s} G'(\ln s) X(ks),
\end{equation}
and
\begin{align}
W(x) ={}& \frac{3 \sin(2x)}{2 x^3}
- \frac{3 \cos(2x)}{x^2} - \frac{3\sin(2x)}{2x}, \nonumber\\
X(x) ={}& \frac{3}{x^2} (\sin x -x\cos x)^2,
\end{align}
such that the power spectrum freezes out after sound horizon crossing $k s \ll 1$.  The presence of the $I_1^2$ term, which is quadratic in $G'$, allows for ${\cal O}(1)$ power spectrum features.  Imposing  a prior constraint of $I_1^2 < 1/2$ then forbids the breakdown of the GSR approximation itself.

Beyond slow-roll, deviations from scale invariance must be introduced in the time domain via $G'$ because the sharper its features, the more the curvature power spectrum ${\cal P}(k)$ rings due to the oscillations in the $W(x)$ and $X(x)$ freezeout kernels.   Arbitrarily sharp but non-oscillatory features in ${\cal P}(k)$ itself are forbidden in single-field inflation.
The advantage of using GSR is that a sharp deviation from scale invariance in time during inflation is consistently treated in the power spectrum.

\section{Datasets and Methodology}
\label{sec:data}

\subsection{GSR Templates}

\begin{figure}
    \centering
    \includegraphics[width=\linewidth]{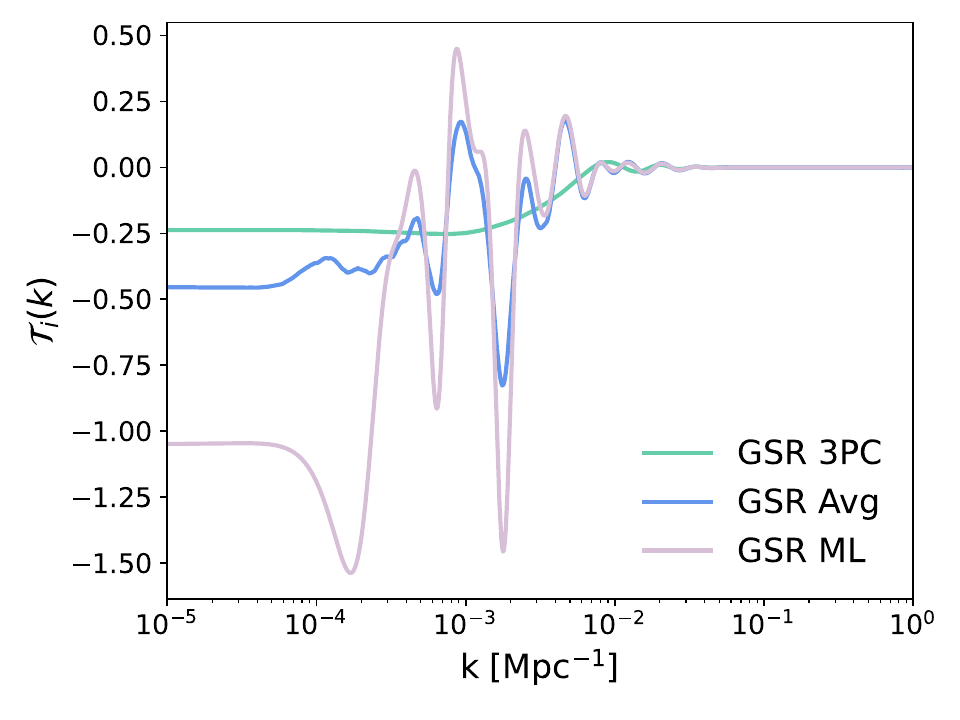}
    \caption{GSR templates for the large-scale curvature suppression favored by the large-angle lowT CMB anisotropy extracted from  Ref.~\cite{Obied:2018qdr}.  The GSR 3PC template contains the 3 most significant principal components and represents a steplike suppression near $k\sim 10^{-3}$ Mpc$^{-1}$, whereas the GSR avg better fits the oscillatory shape of the anomaly while still preserving the suppression.   The GSR ML template overfits low significance fluctuations in the data but is similar in this region where the feature is most significant.}
    \label{fig:templates}
\end{figure}

We begin with the parametrization of Ref.~\cite{Obied:2017tpd}, where the freedom in the GSR source function $\delta G'$ is described by 20 logarithmically spaced spline knots in the range $s=200-20000$\,Mpc.  
While this is more than sufficient to fit large scale features in CMB spectra, it also risks overfitting statistical fluctuations in the data with very low significance variations in $G'$.  We therefore reduce this freedom to a single parameter $A_i$ for each of 3 template forms that are informed by the features in the lowT data and extracted from the analysis of Ref.~\cite{Obied:2018qdr}
\begin{equation}
\label{eq:templates}
\ln \mathcal{P}(k) =\ln  A_s + (n_s-1)\ln \left( \frac{k}{k_0 }\right)+  A_i {\cal T}_i(k),
\end{equation}
where ${\cal T}_i(k)$ are the templates and $i \in \{$3PC,\,avg,\,ML$\}$ as shown in
Fig.~\ref{fig:templates} which we now describe.

The first template ${\cal T}_{3\rm PC}$ comes from the 3 most significant aspects of the deviation from scale invariance identified by the principal component (PC) analysis of Ref.~\cite{Obied:2018qdr}.
It takes the form of a steplike suppression of large scale power. Our 3PC template analysis therefore can be thought of as fixing the location and width of the step while leaving the amplitude free.  Again the advantage of parameterizing it in the GSR time domain rather than a step in the power spectrum is that as the width of the step decreases, the feature in ${\cal P}(k)$ consistently develops the ringing features.   These are visible at $k \sim 10^{-2}$ Mpc$^{-1}$ in Fig.~\ref{fig:templates}.

This steplike template only captures the overall suppression of the curvature power spectrum whereas the $\ell=20-40$ feature in lowT resembles an oscillation around this net suppression, with both local power suppression and enhancement.   To explore the impact of the more detailed form, we also extract the average of the GSR power spectrum posterior deviations from Ref.~\cite{Obied:2018qdr} as second template ${\cal T}_{\rm avg}$.  Notice the prominent oscillations near $k\sim 10^{-3}$\,Mpc$^{-1}$ around the  ${\cal T}_{\rm 3PC}$ steplike form.
In GSR, oscillations in the power spectrum are a natural consequence of sharp inflationary features in $\epsilon_H$ whereas they might appear more ad hoc when parameterizing the power spectrum directly.

Finally we take the maximum likelihood GSR model to define the last template ${\cal T}_{\rm ML}$.  This wildly varying template clearly overfits statistical fluctuations, especially for $k \lesssim 10^{-4}$ Mpc$^{-1}$ where even ${\cal T}_{\rm avg}$ has overfitted small features like the highly suppressed quadrupole.  Nonetheless ${\cal T}_{\rm ML}$ has a very similar form as ${\cal T}_{\rm avg}$ for the $10^{-3}$\,Mpc$^{-1}$ region. We  use ${\cal T}_{\rm ML}$ to test the robustness of inferences on $\tau$, not for assessing the significance of the lowT features.

\subsection{Data}

The goal of this work is to characterize the degree to which inflationary features at large scales in conjunction with BAO data can raise the optical depth within $\Lambda$CDM. Since large scale features are only measured in all-sky data, we first investigate the implications of the Planck primary anisotropy data, and then proceed to add more recent fine scale data from the Atacama Cosmology Telescope (ACT) and the South Pole Telescope (SPT), as well as lensing data from all three, and finally BAO data from DESI: 

\begin{itemize}
    \item Planck: We use the PR3 low-$\ell$ TT likelihood (lowT), the \texttt{SRoll2} low-$\ell$ EE likelihood (lowE) \cite{Delouis:2019bub}, and the \texttt{Plik} high-$\ell$ TTTEEE (lite) \cite{Planck:2018vyg}.  Note that we do not include Planck lensing reconstruction in this combination.
        
    \item CMB: We adopt the latest data from the Atacama Cosmology Telescope (ACT DR6) and the South Pole Telescope (SPT-3G D1), using the lite versions of the likelihood from \texttt{candl}. We combine these with the same Planck likelihoods as above, but use Planck data at $\ell < 1000$ for TT and $\ell < 600$ for TE and EE and ACT data above those multipole values to reduce the covariance between ACT DR6 and Planck, following \cite{ACT:2025fju}. Since SPT and Planck/ACT do not have significant overlap, we use the full multipole range of SPT. We use the combined Planck PR4 and ACT DR6 lensing likelihood, accounting for their cross-correlations and implement SPT-3G MUSE lensing data as an independent dataset, treating cross-correlations as negligible.
    The ``no lowE" CMB combination refers to the combination of these data excluding the lowE likelihood.

    \item BAO: We use the DESI DR2 likelihood 
    based on Table \rev{IV} in  Ref.~\cite{DESI:2025zgx} for $D_V/r_d$ for the lowest redshift bin and $D_M/r_d$, $D_H/r_d$, and their cross-correlation $r_{M,H}$ for all other bins.
Here 
\begin{equation}
    D_M(z_i) = \int_0^{z_i}\frac{dz}{H(z)} \,,
\end{equation}
$D_H(z)=1/H(z)$, 
$D_V(z) = 
[z D_M^2(z) D_H(z)]^{1/3}$ and $r_d$ is the photon-baryon sound horizon at the end of the Compton drag epoch. 
\end{itemize}

\subsection{Methodology}

We compute power spectra with the Boltzmann code \texttt{CLASS} (Cosmic Linear Anisotropy Solving System).\footnote{\url{https://github.com/lesgourg/class_public}}
To implement the GSR templates above, we modify  the initial curvature power spectrum by providing an externally calculated array from Eq.~(\ref{eq:templates}), using
 the built-in functionality within \texttt{CLASS}. \\

We sample the Bayesian posteriors through a Metropolis-Hastings Markov chain Monte Carlo (MCMC) analysis with \texttt{cobaya},\footnote{\url{https://github.com/CobayaSampler/cobaya}}
and  analyze the posteriors with  the \texttt{Getdist} package.\footnote{\url{https://getdist.readthedocs.io}}
All our chains are converged to a Gelman-Rubin convergence statistic of $R-1 < 0.01$. 
When quoting the maximum allowed optical depth  $\taumax$ we use the one-sided 95\% CL upper limit from the posterior.  For the no lowE analysis only, we also quote the minimum optical depth as $\tau_{\rm min}$, the one-sided 95\% CL lower limit from the posterior.\\

Additionally, we use \texttt{cobaya}'s \texttt{bobyqa} algorithm to find best-fits for each model and data combination, complementing our Bayesian posteriors by showing the tradeoffs between different datasets. We report these $\Delta\chi^2\equiv-2\Delta\ln {\cal L}$ values
relative to the $\Lambda$CDM (no GSR template) best-fit for the given data combination.  We use the sign convention that negative \dchisq\ values represent a higher likelihood value ${\cal L}$. \\

We use flat bounded priors for the template amplitudes $A_i \in [-2, 3]$. For the other parameters we take flat priors on  $\ln(10^{10} A_\mathrm{s}) \in [1, 4]$; $n_s \in [0.8,1.2]$; $\tau \in [0.004, 0.8]$;  cold dark matter density $\Omega_c h^2 \in [0.001, 0.99]$; baryon density $\Omega_b h^2 \in [0.005, 0.1]$; and Hubble constant $H_0 \in [20,100]$ km\,s$^{-1}$\,Mpc$^{-1}$.  These priors are all uninformative compared with the posteriors from the chosen data combinations.   We fix the sum of the neutrino masses to their minimal value $\sum m_\nu =0.06$\,eV.

\section{Raising $\tau$ by lowering ${\cal P}$}
\label{sec:planck}

At large scales, the EE power spectrum of the CMB polarization contains information about reionization, specifically through the “reionization bump” at $\ell\lesssim20$.  This lowE bump is generated from the local quadrupole anisotropy of the CMB intensity, generated by the Sachs-Wolfe effect, scattering off of free electrons during reionization. 
Since quadrupole anisotropy is the source of polarization, the multipoles of the bump correspond to the comoving horizon scale at scattering.  
The shape of the bump therefore has a weak dependence on the detailed ionization history \cite{Hu:2003gh} and, more importantly for this work, has a hidden dependence on the curvature power spectrum 
\begin{equation}
C_\ell^{EE} \propto \tau^2 {\cal P}(k\sim \ell/D_M(z_{\rm rei})),
\end{equation}
where in the standard reionization history for  the effective redshift $z_{\rm rei}$, $k \sim 10^{-4} - 10^{-3}$ Mpc$^{-1}$ for the reionization bump (see \cite{Mortonson:2009qv}, Fig.~3).

\begin{figure}
    \centering
    \includegraphics[width=\linewidth]{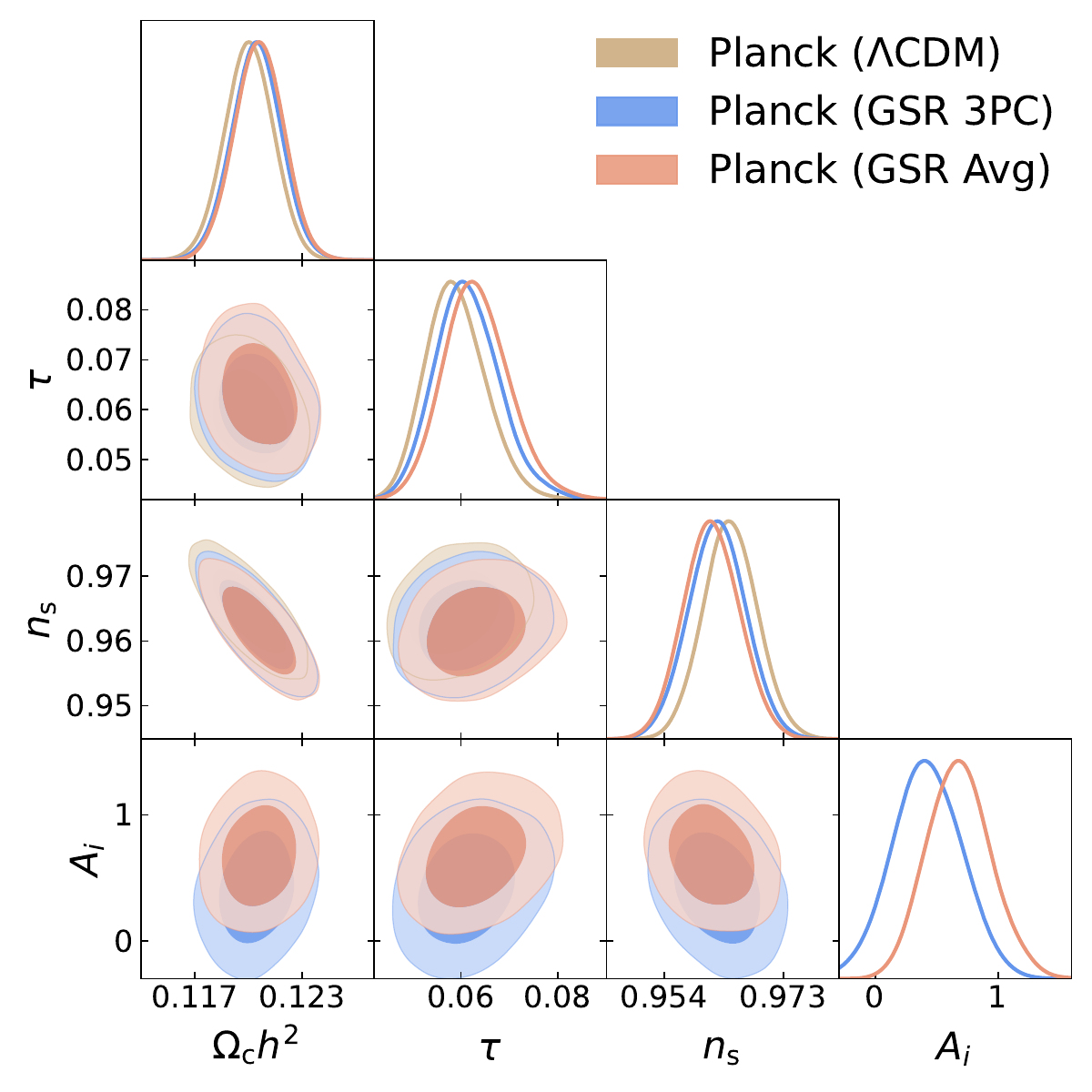}
    \hfill
    \caption{Posterior constraints with Planck primaries.  The $\Lambda$CDM result without large scale curvature features, where $\tau=0.0589^{+0.0055}_{-0.0067}$, is compared with posteriors where the amplitudes $A_i \in \{$3PC, avg$\}$ are marginalized.  The correlation between lowering the curvature by raising $A_i$ and raising $\tau$ leads to  $(A_{\rm 3PC} = 0.42\pm 0.29, \tau=0.0613^{+0.0061}_{-0.0071})$ and
    $(A_{\rm avg} = 0.68^{+0.25}_{-0.28},\tau = 0.0632^{+0.0062}_{-0.0073})$.  The better the fit to the lowT feature, the higher $\tau$ becomes.
}
    \label{fig:planck3pc}
\end{figure}

\begin{figure}
    \centering
    \includegraphics[width=\linewidth]{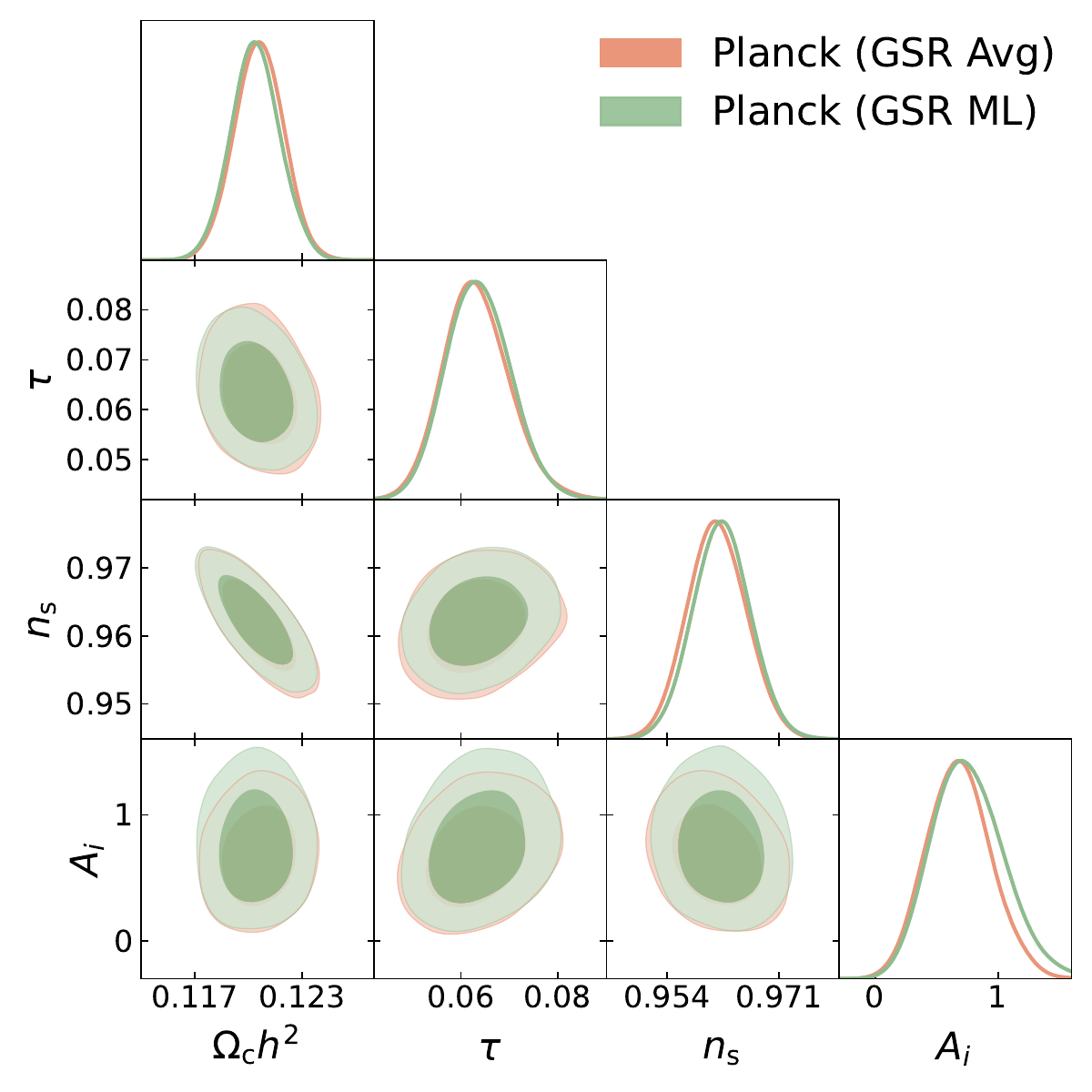}
    \hfill
    \caption{Planck posteriors between the GSR avg and ML templates demonstrating the robustness of the $\tau$ constraint.   Further (over)fitting of the large angle temperature features with this maximum likelihood form raises the amplitude of the curvature feature to 
    $A_{\rm ML} = 0.76^{+0.26}_{-0.32}$ but leaves the optical depth posteriors essentially unchanged at
 $\tau = 0.0636\pm 0.0067$.}
    \label{fig:planckML}
\end{figure}

For slow-roll curvature power spectra, extrapolation of the high-$k$ constraints to low-$k$ through the scalar tilt $n_s$ fixes the value of ${\cal P}$ here and the lowE measurements themselves then determine $\tau$.  However this is not true of models with 
large scale curvature features
\cite{Mortonson:2009xk}. 
In particular for models that lower the large scale curvature power by some $\Delta {\cal P}$, we expect an increase of the optical depth of 
\begin{equation}
\frac{\Delta \tau}{\tau} \approx -
\frac{1}{2} \frac{\Delta {\cal P}}{{\cal P}}.
\end{equation}
For example with the steplike 3PC GSR template in Fig.~\ref{fig:templates} and $A_{3{\rm PC}}=1$, $\Delta {\cal P}/{\cal P}\approx -0.24$ and one would expect $\Delta\tau/\tau \approx 0.12$ if all other parameters were held fixed.

These expectations are borne out by the 3PC analysis of the Planck primary data in Fig.~\ref{fig:planck3pc} (blue contours).  Note the correlation between $\tau$ and $A_{\rm 3PC}$ that shifts the mean value of $\tau$ higher compared with the $\Lambda$CDM result (red contours). 
  The posterior constraints give $A_{3{\rm PC}}=0.42\pm 0.29$ 
 which raises the mean value over $\Lambda$CDM by 
$\Delta\tau/\tau= 0.041$ and the one-sided 95\% upper limit to $\taumax=0.0728$
(see Tab.~\ref{tab:planck}).  
 On the other hand, the preferred amplitude of the template is reduced from unity, leaving only a nominal $1.4\sigma$ preference for a finite amplitude.  Correspondingly the best-fit in Tab.~\ref{tab:planck} only improves by $\Delta\chi^2\approx -2$ and the lowT likelihood itself by $\Delta\chi^2_{\rm lowT}\approx -3$ for the one extra parameter.

This reduced amplitude and lowT improvement is in part due to the fixed steplike form of the template, which does not optimally fit the lowT feature.   
In Fig.~\ref{fig:planck3pc} (brown contours) we show the results for the average template amplitude $A_{\rm avg}$ instead.   For the Planck primary analysis $A_{\rm avg}= 0.68^{+0.25}_{-0.28}$ is larger with a best-fit improvement of $\Delta\chi^2 \approx -7$ (Tab.\ \ref{tab:planck}),  the majority of which comes from the lowT likelihood.
Likewise the implications for raising the optical depth are stronger with  
$\tau = 0.0632^{+0.0062}_{-0.0073}$
and $\taumax=0.0751$.   Note that because the template is no longer a uniform step lower in curvature, the best-fit amplitude does require a compromise between the lowE and lowT data.   Furthermore, the oscillations in the template at $k>10^{-3}$ Mpc$^{-1}$ also play a role in improving the TTTEEE high-$\ell$ likelihood, with $\Delta \chi^2_{\rm TTTEEE} \sim -3.3$ in Tab. \ref{tab:planck}. Overall the average template allows a better fit to the data with a higher $\tau$ than in $\Lambda$CDM. 

In summary, the marginal posteriors of \{\lcdm, 3pc, avg\} in Fig.~\ref{fig:planck3pc} show the sequence of increasing $\tau$ with cases that better fit the aspects of Planck primaries other than the lowE data. Note that these results should not be taken as evidence for the lowT anomaly, since the templates are obtained \textit{a posteriori}. Instead, they show that a larger $\tau$ (and the corresponding changes in fit to different data) is a prediction of interpreting the lowT anomaly as a physical feature.  Remarkably, the resulting $\taumax \approx 0.075$ already overlaps the preferred CMB+BAO range that excludes lowE, as discussed in Sec.~\ref{sec:BAO}.

\renewcommand{\arraystretch}{1.2}

\begin{table*}
    \centering
    \begin{tabular}{|l|c|c|c|c|c|c|c|c|c|}
    \hline
    \multicolumn{10}{|c|}{\textbf{Planck}} \\
    \hline
    \textbf{Model} 
    & $\boldsymbol{\Delta\chi^2}$ 
    & $\boldsymbol{\Delta\chi^2_{\rm lowT}}$ 
    & $\boldsymbol{\Delta\chi^2_{\rm lowE}}$ 
    & $\boldsymbol{\Delta\chi^2_{\rm TTTEEE}}$ 
    & $\boldsymbol{A_i}$ 
    & $\boldsymbol{\tau}$ 
    & $\boldsymbol{n_s}$
    & $\boldsymbol{\Omega_c h^2}$
    & $\boldsymbol{\tau_{\rm max}}$
    \\
    \hline
    $\Lambda$CDM & 0 & 0 & 0 & 0 & 0 & 0.0584 & 0.9646 & 0.12008 & 0.0696 \\ 
    GSR 3PC & -2.01 & -3.00 & 0.27 & 0.72 & 0.412 & 0.059 & 0.9627 & 0.12043 & 0.0728 \\
    GSR avg & -6.99 & -5.76 & 2.11 & -3.34 & 0.681 & 0.0629 & 0.9620 & 0.12038 & 0.0751 \\
    GSR ML & -8.33 & -7.08 & 2.76 & -4.01 & 0.652 & 0.0620 & 0.9622 & 0.12034 & 0.0749\\
    \hline
\end{tabular}
    \caption{Best-fit parameter values, $\Delta\chi^2$ breakdown relative to $\Lambda$CDM, and $95\%$ upper limit $\taumax$ for the Planck primary analysis. The GSR templates suppress power in ${\cal P}(k)$ at large scales, increasing the inferred value of $\tau$. While the GSR templates improve the fit to the lowT data by their {\it a-posteriori} construction, the average and maximum likelihood ones also improve the fit to the high-$\ell$ TTTEEE data.  }
    \label{tab:planck}
\end{table*}

Finally to check the robustness of the upper limits on $\tau$, we analyze the maximum likelihood template in Fig.~\ref{fig:planckML}.  The slightly larger amplitude $A_ {\rm ML} = 0.76^{+0.26}_{-0.28}$ is accompanied by a negligible additional change in $\tau=0.0636\pm 0.0067$ and $\taumax=0.0749$.  Note the close match of $\tau$ posteriors for the average and ML templates in Fig.~\ref{fig:planckML}.   We conclude that the results for how high $\tau$ can be when interpreting the lowT feature as inflationary rather than a statistical anomaly are robust.

As an aside, note that the amplitude $A_{\rm ML}<1$ at $\sim 1\sigma$ is still smaller than implied by the Planck 2015 analysis where the best fit $A_{\rm ML}=1$.  Likewise, the further improvement in the lowT likelihood only increases to $\Delta\chi^2_{\rm lowT}\approx -7$ with a total of $\Delta\chi^2 \approx -8.3$.
The change between Planck 2015, from which the template amplitudes are normalized, and our current analysis is mainly the intervening improvements in the lowE data themselves due to  Planck's High Frequency Instrument (HFI). These shifted the $\Lambda$CDM value for $\tau$ lower, from $\tau=0.079\pm 0.01$ \cite{Planck:2015fie} to the current values.  

A smaller $\tau$ makes the lowT feature less significant since even without a feature, the Sachs-Wolfe plateau is lower compared with the acoustic peaks 
(see \cite{Obied:2018qdr} their Fig.~11).  It then reduces the need to introduce a feature in the large scale curvature to accommodate the lowT data.  
We therefore attribute our $\sim 1\sigma$ lowered value of $A_{\rm ML}$ at least in part to the change in the lowE polarization data between the 2015 and final Planck release.  We reserve for future work a full study of implications for the curvature power spectrum itself, using  different combinations of new data sets,  as opposed to the optical depth treated here.\footnote{Preliminary results indicate that the implications for $\taumax$ for our cases do not change significantly.}

\begin{figure}
    \centering
    \includegraphics[width=\linewidth]{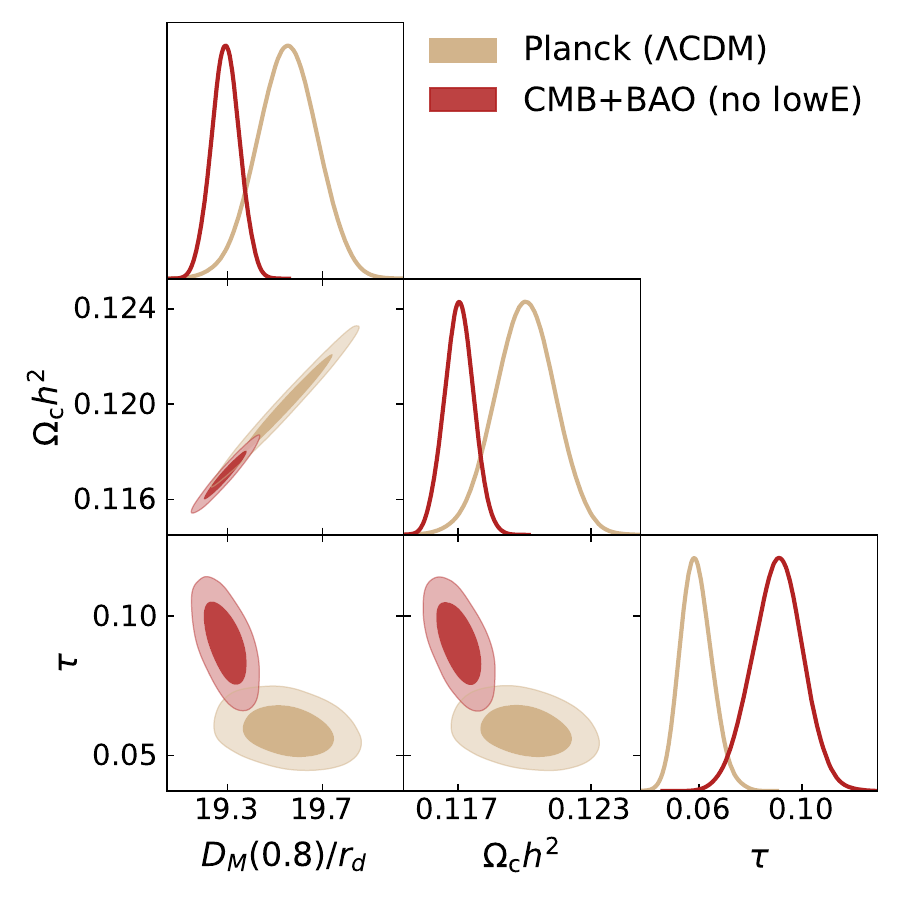}
    \caption{Posteriors for the parameters relevant for CMB+BAO tension.   BAO favor a lower distance $D_M(z\sim 0.8)/r_d$ and indicate a small value for $\Omega_c h^2$ in $\Lambda$CDM.  With constraints from lowE, the two are in tension due to the anticorrelation between $\tau$ and $\Omega_c h^2$.  In $\Lambda$CDM with no lowE data this tension is resolved by a higher $\tau =0.0904 \pm 0.0098$ and a lower $\Omega_c h^2 =0.11704\pm 0.00067$.  CMB+BAO tension then is shifted to the incompatibility of the $\tau$ constraints: $\taumax=0.0696$ for Planck ($\Lambda$CDM) and $\tau_{\rm min}=0.074$ for CMB+BAO (no lowE).}
    \label{fig:dm_rd_tau}
\end{figure}

\section{Raising $\tau$ with BAO}
\label{sec:BAO}

Although the increase in $\tau$ from inflationary lowT features is modest when considering Planck primaries alone, the upwards shift is potentially more important because the combination of CMB and BAO measurements also favor high $\tau$.  Moreover with the CMB+BAO combination, the maximum $\tau$ rises even further with potential implications for the high-redshift Universe.

Under $\Lambda$CDM, CMB and BAO data are $2-3 \sigma$ inconsistent \cite{DESI:2025zgx}, with BAO preferring  a larger value of  the relative distance between $z\sim 0.8$ and $z_*\sim 1100$, $[D_M(z_*)-D_M(0.8)]/r_d$ or equivalently a 1-2\% lower $D_M(0.8)/r_d$ when the angular scale of the CMB sound horizon  is pinned to the CMB measurements, which also fixes $D_M(z_*)/r_d$
\cite{Liu:2025bss}.  The shape of CMB primary anisotropy and lensing amplitude determine the value of $\Omega_c h^2$ which calibrates $r_d$ and controls relative distance at high $z$. If it were not for the lowE data, these shifts could be accommodated by lowering $\Omega_c h^2$ through raising $\tau$ to $\sim 0.09$ \cite{Jhaveri:2025neg,Sailer:2025lxj}.  Viewed this way, CMB+BAO tension can be rephrased as a tension between lowE direct measurements of $\tau$ and the $\Lambda$CDM inference of $\tau$ from the rest of the CMB data and BAO.

Fig.~\ref{fig:dm_rd_tau} shows this solution by explicitly dropping the lowE data when adding the BAO data.   Notice the anticorrelation between $\tau$ and  $\Omega_c h^2$ that allows the shift downward in $D_M(0.8)/r_d$ from the range preferred by the full Planck data in $\Lambda$CDM.

On the other hand as seen in Fig.~\ref{fig:dm_rd_tau}, in $\Lambda$CDM this solution would be excluded by the lowE Planck data taken at face value which confine $\tau$ to lower values and force the compromise that exhibits the original CMB+BAO tension in $\Lambda$CDM.
The one-sided 95\% {\it lower} limit for the no lowE results is $\tau_{\rm min}= 0.074$ which is inconsistent with the Planck $\Lambda$CDM $\taumax= 0.0696$ from Tab.~\ref{tab:planck}.
This tension then motivates the exploration of models beyond this slow-roll $\Lambda$CDM and standard reionization to reconcile these inferences for $\tau$ (see \cite{Jhaveri:2025neg} for changes to the reionization history).

This shift in $\Omega_c h^2$ downwards and $\tau$ upwards, and the tradeoff in other parameters required, is reminiscent of the well-studied  but larger changes in cosmological parameters between the full Planck data set and WMAP9 \cite{WMAP:2012nax} where $\Omega_c h^2=0.1138 \pm 0.0045$ was even lower (and 
$\tau=0.089 \pm 0.014$)
or Planck TT $\ell<1000$ 
\cite{Planck:2016tof}.  
In the data, these were driven by the smoothness of the acoustic peaks or ``oscillatory residuals" at $\ell>1000$ and the interpretation of the lowT anomaly within $\Lambda$CDM
\cite{Planck:2016tof}.
Lowering $\Omega_c h^2$ in $\Lambda$CDM creates tension in both the primary CMB anisotropy and, if too far, CMB lensing reconstruction.    
On its own, a smaller $\Omega_c h^2$ changes the shape of the primary anisotropy by raising acoustic driving effects and reducing lensing, both of which leave oscillatory residuals that imply sharper acoustic peaks at $\ell>1000$ than observed. 

This sharpness problem can be removed by artificially amplifying lensing by rescaling its amplitude $A_L$ for primaries. To reach the more extreme levels of $\Omega_c h^2$ truly favored by BAO shown in Fig.~\ref{fig:dm_rd_tau}, the lensing reconstruction amplitude must also be similarly rescaled \cite{Green:2024xbb,SPT-3G:2024atg}, implying that lensing reconstruction also disfavors the very low values of $\Omega_c h^2=0.11704 \pm 0.00067$ now preferred by BAO (see Fig.~\ref{fig:dm_rd_tau}).  Lowering the neutrino mass to formally negative values also has effects on the primary anisotropy and lensing reconstruction similar to $A_L$, but neither option is physical \cite{Green:2024xbb,Jhaveri:2025neg,Loverde:2024nfi}.

Without lowE data, raising $\tau$ presents a physical alternative to reconciling CMB+BAO in $\Lambda$CDM \cite{Jhaveri:2025neg,Sailer:2025lxj}.  
Raising $\tau$ lowers $\Omega_c h^2$ to preserve the shape of $C_\ell$ while also raising $A_s$ at fixed $A_s e^{-2\tau}$ to enhance lensing.  Both effects counter the overly sharp peaks while simultaneously preventing the underprediction of lensing.
The best-fit no lowE model is given in Tab.~\ref{tab:cmb_bao} 
with its low value of $\Omega_c h^2\approx 0.117$ and high value of $\tau\approx 0.09$ improving the 
$\chi^2$ of both high-$\ell$ temperature and polarization (highTE).
Even with lowE data included in the combination of CMB+BAO data, the best-fit values in $\Lambda$CDM are shifted downwards to $\Omega_c h^2\approx 0.118$ and upwards to $\tau \approx 0.0685$ from their Planck best-fit values of $\Omega_c h^2\approx 0.12$ and $\tau\approx 0.0584$ in Tab.~\ref{tab:planck}.  These shifts represent a compromise that makes the CMB fit in $\Lambda$CDM worse overall compared with the no lowE case.

The interpretation of the lowT feature in the data also impacts these shifts \cite{Planck:2016tof,Obied:2017tpd}.  
In $\Lambda$CDM with only TT at $\ell<1000$, the lowT feature can be interpreted as a larger $n_s$, with  similar compensations of lower $\Omega_c h^2$, higher $H_0$, higher $\Omega_b h^2$ in the direction toward the WMAP9 cosmology \cite{WMAP:2012nax}. With the full Planck range this is no longer possible and the lowT feature 
becomes a more significant statistical anomaly as a consequence.

\begin{figure}
    \centering
    \includegraphics[width=\linewidth]{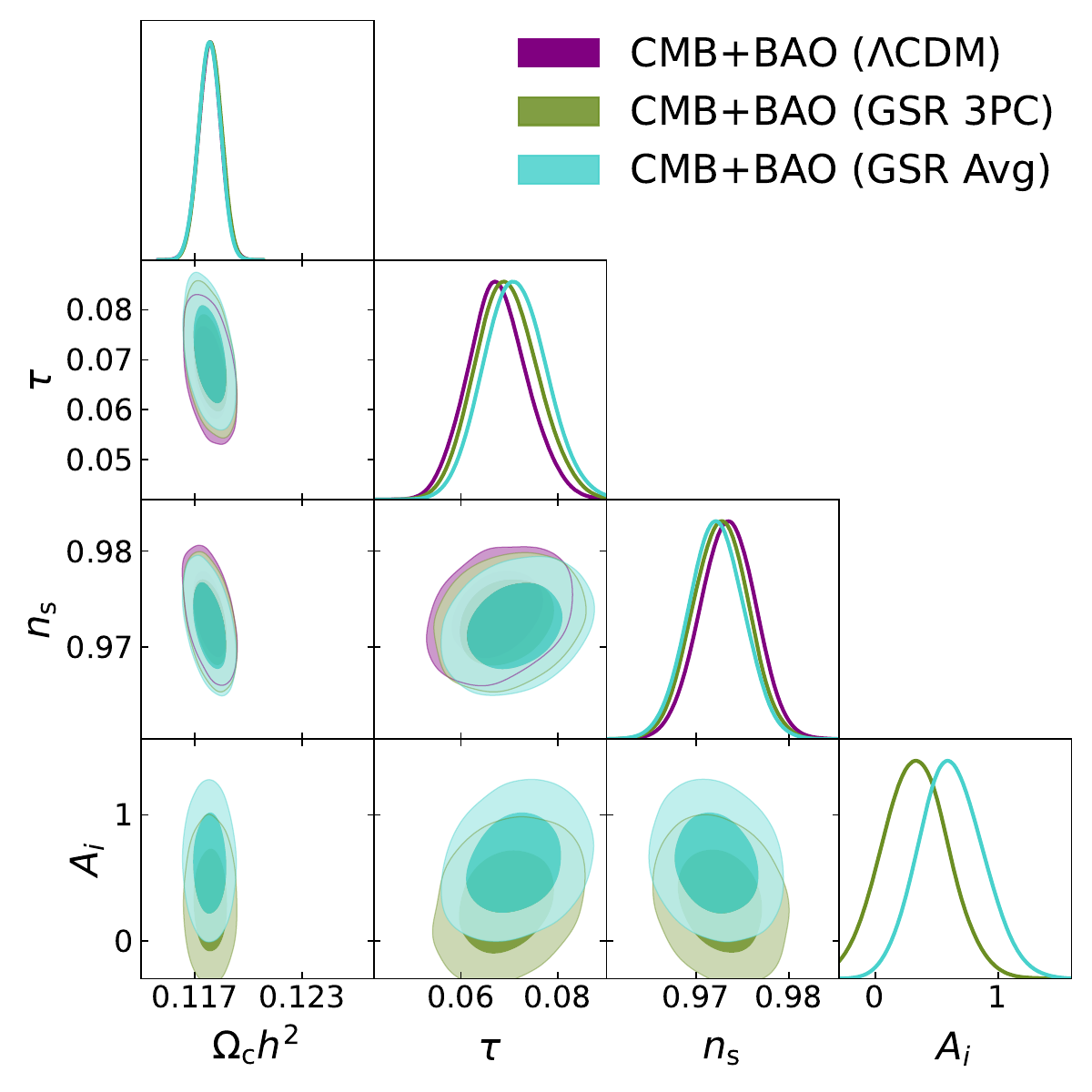}
    \hfill
    \caption{Posterior constraints with CMB+BAO data.  
    The inclusion of BAO data systematically shift $\tau$ higher and $\Omega_c h^2$ lower as compared with similar Planck primary analysis in Fig.~\ref{fig:planck3pc}. Compared with $\Lambda$CDM where $\tau = 0.0677^{+0.0059}_{-0.0067}$, the cases with large scale curvature features give $(A_{\rm 3PC} = 0.32\pm 0.27;\tau = 0.0697^{+0.0061}_{-0.0068})$ and
    $(A_{\rm avg} = 0.62^{+0.24}_{-0.28},  \tau = 0.0712\pm 0.0064)$.  The template cases further raise $\tau$ but do not further lower $\Omega_c h^2$ from their $\Lambda$CDM values.}
    \label{fig:cmb_bao_3pc}
\end{figure}

\begin{figure}
    \centering
    \hfill
    \includegraphics[width=\linewidth]{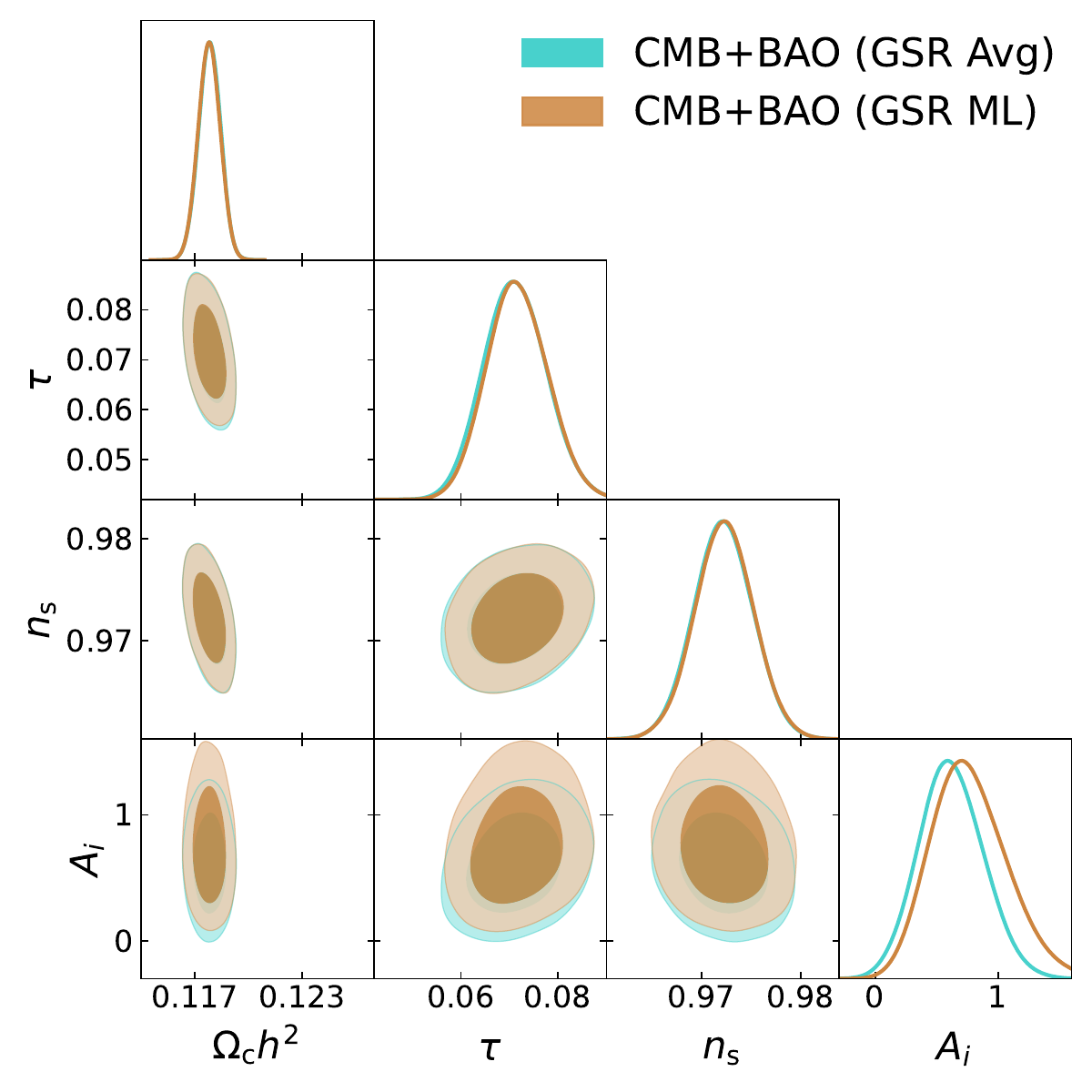}
    \caption{CMB+BAO posteriors between the GSR avg and ML templates, demonstrating the robustness of $\tau$.  As with Fig.~\ref{fig:planckML} for the Planck primary analysis, using the GSR ML templates raises  $A_{\rm ML} = 0.77^{+0.26}_{-0.35}$ but leaves $\tau = 0.0715^{+0.0059}_{-0.0069}$ relatively unchanged from the avg template results.}
    \label{fig:cmb_bao_ML}
\end{figure}

With
the GSR ability to change the lowT spectrum through the curvature, this statistical anomaly is reduced, producing a better fit to the CMB at the expense of additional parameters. The other $\Lambda$CDM parameters shift slightly to lower $n_s$ and larger $\Omega_c h^2$ since the intrinsic height of the acoustic peaks relative to slow-roll expectations is reduced to compensate for the reinterpretation of the lowT feature as an inflationary deficit \cite{Obied:2017tpd}.
This can be seen for the Planck primaries in 
Fig.~\ref{fig:planck3pc} and interestingly the same trends exist when replacing Planck TT at $\ell>1000$ with ACT and SPT in Fig.~\ref{fig:cmb_bao_3pc} although the correlated shift of $n_s$ is slightly weakened as compared with Fig.~\ref{fig:planck3pc} due to the extended multipole range.
Furthermore with CMB+BAO, the ability of the GSR feature to slightly increase $\Omega_c h^2$ is no longer allowed. The best-fit values of $\Omega_c h^2$ are only very weakly sensitive to the variations in Tab.~\ref{tab:cmb_bao} and its posterior is almost decorrelated with the template amplitudes $A_i$ in Fig.~\ref{fig:cmb_bao_3pc}.

Nonetheless adding BAO data does increase $\tau$ further for the template cases as well, due mainly to a better fit on the CMB side of the combination.
With BAO data, $\tau$ is further increased to $\taumax=0.0805$ for the 3PC template and $\taumax=0.0820$ for the average template.
The impact of higher $\tau$ is mainly to improve the compatibility of these similar low $\Omega_c h^2$ values with CMB data rather than to drive $\Omega_c h^2$ even lower to better fit the BAO data themselves as is possible with the no lowE case.  
This can be seen in Tab.~\ref{tab:cmb_bao} where the total improvement is dominated by the CMB side, especially lowT and highTE.   The total improvement for the 3PC case is not significant but for $A_{\rm avg}$ reaches $\Delta\chi^2 \approx -5.77$ vs $\Lambda$CDM for the one extra parameter, albeit one that is introduced {\it a posteriori} to fit the temperature anomaly.  Here $-3.64$ comes from highTE and lensing. 
On the other hand raising $\tau$ by lowering large scale power is not as effective in resolving CMB+BAO tension as other means of raising $\tau$, e.g.~through unknown data systematics or non-standard reionization histories \cite{Jhaveri:2025neg,Tan:2025obi}, which can lower $\Omega_c h^2$ further.

Finally  in Fig.~\ref{fig:cmb_bao_ML} we  show that results for $\tau$ when marginalizing  ${A}_{\rm ML}$  are again very similar to that of marginalizing $A_{\rm avg}$.  Here $\taumax=0.0821$ and the $\tau$ posteriors  are nearly identical.  We conclude that even with CMB+BAO data, the implications for $\tau$ are robust to further overfitting of the lowT feature.

\renewcommand{\arraystretch}{1.2}

\begin{table*}
    \centering

\begin{tabular}{|l|c|c|c|c|c|c|c|c|c|c|c|}
    \hline
    \multicolumn{12}{|c|}{\textbf{CMB + BAO}} \\
    \hline
    \textbf{Model} 
    & $\boldsymbol{\Delta\chi^2}$ 
    & $\boldsymbol{\Delta\chi^2_{\rm lowT}}$ 
    & $\boldsymbol{\Delta\chi^2_{\rm lowE}}$ 
    & $\boldsymbol{\Delta\chi^2_{\rm highTE}}$ 
    & $\boldsymbol{\Delta\chi^2_{\rm lens}}$ 
    & $\boldsymbol{\Delta\chi^2_{\rm BAO}}$ 
    & $\boldsymbol{A_i}$ 
    & $\boldsymbol{\tau}$ 
    & $\boldsymbol{n_s}$
    & $\boldsymbol{\Omega_c}h^2$ 
    & $\boldsymbol{\tau_{\rm max}}$ \\
    \hline
    $\Lambda$CDM & 0 & 0 & 0 & 0 & 0 & 0 & 0 & 0.0685 & 0.9738 & 0.11775 & 0.0782 \\ 
    3PC & -1.15 & -1.99 & 0.50 & 0.70 & -0.35 & -0.01 & 0.357 & 0.0708 & 0.9728 & 0.11776 & 0.0805 \\
    GSR avg & -5.77 & -4.05 & 1.59 & -3.13 & -0.51 & 0.33 & 0.566 & 0.0720 & 0.9726 & 0.11782 & 0.0820 \\
    GSR ML & -7.92 & -5.38 & 1.67 & -3.72 & -0.12 & -0.36 & 0.691 & 0.0704 & 0.9728 & 0.11765 & 0.0821 \\
    No lowE & -4.64 & 0.31 & - & -1.33 & -1.28 & -2.34 & 0 & 0.0892 & 0.976 & 0.11699 & 0.1063 \\
    \hline
\end{tabular}
    \caption{Best-fit parameter values, $\Delta\chi^2$ breakdown relative to $\Lambda$CDM and 95\% upper limit $\taumax$ for the CMB+BAO analysis. In all cases, tension between CMB and BAO data drives $\taumax$ higher to accommodate. 
    The template models improve the fit to CMB highTE and lensing (in addition to the lowT). However, they leave $\Omega_c h^2$ and hence $\Delta\chi^2_{\rm BAO}$ nearly the same, unlike the no lowE case. This is due to small shifts in $n_s$ and $\Omega_ch^2$ when interpreting the lowT anomaly as an inflationary feature (see Fig.~\ref{fig:planck3pc} and Sec.~\ref{sec:BAO} for details). Here the no lowE total $\Delta\chi^2$ is the sum of the remaining components relative to those values in $\Lambda$CDM (where the fit is to all CMB+BAO).}
    \label{tab:cmb_bao}
\end{table*}

\begin{figure}
    \centering
    \includegraphics[width=\linewidth]{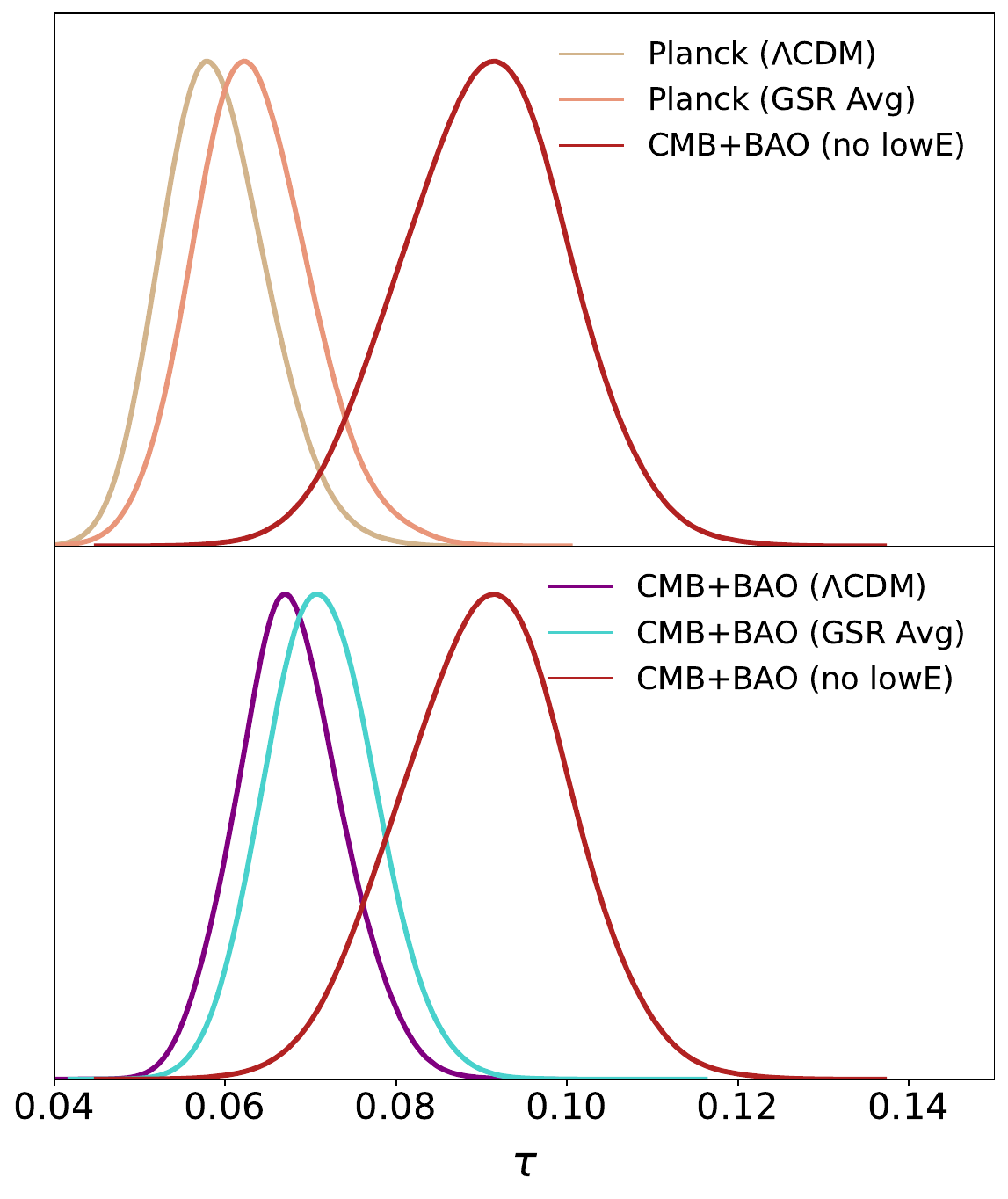}
    \caption{Comparison of the $\tau$ posterior distributions for the most relevant cases. In both panels, the CMB+BAO (no lowE) curve represents the range  needed to bring CMB and BAO data into consistency, with $\tau_{\rm min}=0.074$. \textit{Top panel}: With  Planck primaries data including lowE, $\Lambda$CDM is in tension with $\tau_{\rm max}=0.0696$  but the 95\% interval overlap is restored once the average GSR template is marginalized where $\taumax =0.0751$.  \textit{Bottom panel}: CMB+BAO analysis in $\Lambda$CDM produces $\tau$ constraints that split the tension whereas with the average GSR template they settle into this overlap regime and imply $\taumax=0.0820$. }
    \label{fig:tau}
\end{figure}

Our main results for $\tau$ are summarized in Fig. \ref{fig:tau}. In both panels, we show the “target” of the no lowE analysis that can bring CMB and BAO data fully into consistency. The shift between \lcdm and the GSR $A_{\rm avg}$ marginalized
curves in the top panel reflect what Planck primaries alone imply.  We can see that with GSR features in the large scale curvature $\tau$ posteriors with $\taumax=0.075$ have overlap with the target $\tau_{\rm min}=0.074$ whereas in $\Lambda$CDM they do not.  In the bottom panel we show that when BAO data are included, this overlap region becomes the preferred solution with $\tau=0.0712\pm 0.0064$ and a larger $\taumax=0.082$.

\section{Discussion}
\label{sec:discussion}

In this work, we have explored the freedom provided by large-scale inflationary features in the lowT data to raise the value of $\tau$ with and without the additional boost provided by CMB+BAO tension in $\Lambda$CDM.   These features are produced in the generalized slow-roll formalism to consistently capture the oscillatory effect of transient changes in the slow-roll parameter $\epsilon_H$.  

We adopt templates based on existing fits to the lowT Planck 2015 data \cite{Obied:2018qdr} which extract the most significant features: a steplike suppression of large scale power and an oscillatory modulation around it that accounts for the most significant feature around $\ell=20-40$.  We then marginalize over the amplitude of these templates with a single parameter to address their impact on $\tau$. 
For the Planck primary data alone, we find that the 95\% upper limit for the most representative average template increases from $\taumax=0.0696$ for $\Lambda$CDM to $\taumax  = 0.075$ with a best-fit that reduces $\Delta\chi^2\approx -7$.\footnote{We do not address the statistical significance of the lowT anomaly itself as these template forms are fixed {\it a posteriori}. The single-parameter posterior significance should not be conflated with feature significance due to the look-elsewhere effect.  Instead we address the implications for raising $\tau$ when the existing lowT anomaly is interpreted physically rather than as a statistical fluke.  In this view, a larger $\tau$ is a prediction of such an interpretation which is testable by independent measures.}

A higher $\tau$ is also preferred when BAO data are included due to the CMB+BAO tension in $\Lambda$CDM.  This tension is driven by the low value of the cold dark matter density $\Omega_c h^2$ preferred by BAO, which shifts the cosmological parameters partway back to their pre-Planck WMAP9 values.   In $\Lambda$CDM, the problem in the CMB imposed by the overly sharp acoustic peaks and low lensing power that results from the shifts can be resolved by larger $\tau \sim 0.09$ if the lowE data are dropped \cite{Jhaveri:2025neg, Sailer:2025lxj}.
The CMB+BAO tension in $\Lambda$CDM can then be recast as the incompatibility between the low $\tau_{\rm max}=0.0696$ value inferred from lowE and the high $\tau_{\rm min}=0.074$ value inferred from the rest of the CMB data, including ACT, SPT as well as lensing reconstruction, and BAO.  The GSR marginalized results bring these values back into better compatibility. 

For the implications for $\tau$ itself, we find that marginalizing the average GSR template brings the maximum value to  $\taumax = 0.082$ with the inclusion of the full CMB+BAO data.   This improves the overall fit over $\Lambda$CDM by $\Delta\chi^2 \sim -6$, largely driven by the lowT and high-$\ell$ TTTEEE CMB data, as opposed to the BAO data directly since $\Omega_c h^2$ is not lowered further.   We emphasize that although this does not address the significance of the lowT feature itself, the ability to raise $\tau$ comes as a consequence of interpreting it as a physical effect rather than a statistical fluke.

Direct high-redshift observations may also test this higher-$\tau$ prediction of the lowT anomaly and more generally, the use of $\tau$ in understanding CMB+BAO tension.
JWST measurements of high-redshift galaxies 
\cite{2024ApJ...969L...2F,2022ApJ...938L..15C,2023arXiv230602465E,2023ApJS..265....5H} may  already be providing motivation for earlier formation and hence earlier reionization (e.g.~\cite{Munoz:2024fas,2025Natur.639..897W}).   Conversely, constraints on the duration of reionization from the kinetic SZ effect and Lyman-$\alpha$ forest measurements limit the ability of standard reionization models to raise the optical depth to the BAO-tension resolving $\tau\sim 0.09$ level \cite{Cain:2025usc,Garcia-Gallego:2025jrn} but do not yet exclude these lowT feature-driven levels.

 We also find that while raising $\tau$ to such levels is robust to variants on the template that fits lowT, including the overfitting of the maximum likelihood template, the implications for the curvature power spectrum itself have changed somewhat between Planck 2015 and the current CMB data despite the lowT data anomaly remaining largely unchanged.   This occurs mainly due to the improvement in the lowE data refining the interpretation of the large scale power deficit and the extended multipole range of ACT and SPT refining the underlying slow-roll tilt.   We reserve a full study of the implications for the curvature power spectrum as well as the added ability to change the reionization history beyond the standard form for a future study.
\vfill
\acknowledgments 
We thank Georges Obied for help in extracting GSR templates, Tanvi Karwal for work during the initial stages of this project on external power spectra, and Tom Crawford for useful discussions.
T.J. and W.H. are supported by U.S.\ Dept.\ of Energy contract DE-SC0009924 and the Simons Foundation; T.J. is additionally  supported by the National Science Foundation through awards OPP-1852617 and OPP-233248 under the South Pole Telescope Program.
Computing resources were provided by the  University of Chicago  Research
Computing Center through the Kavli Institute for Cosmological Physics.

 \vfill

\bibliographystyle{apsrev4-2}
\bibliography{references}

\end{document}